# Giant momentum-dependent spin splitting in centrosymmetric low Z antiferromagnets

Lin-Ding Yuan[1], Zhi Wang[1*], Jun-Wei Luo[2], Emmanuel I. Rashba[3], Alex Zunger[1*]

[1]Energy Institute, University of Colorado, Boulder, CO 80309, USA
[2]State Key Laboratory for Superlattices and Microstructures, Institute of Semiconductors, Chinese Academy of Sciences, Beijing 100083, China.
[3]Department of Physics Harvard University, Cambridge, Massachusetts 02138, USA

**Abstract**

The energy vs. crystal momentum E($k$) diagram for a solid (band structure) constitutes the road map for navigating its optical, magnetic, and transport properties. By selecting crystals with specific atom types, composition and symmetries, one could design a target band structure and thus desired properties. A particularly attractive outcome would be to design energy bands that are split into spin components with a momentum-dependent splitting, as envisioned by Pekar and Rashba [Zh. Eksperim. i Teor. Fiz. 47 (1964)], enabling spintronic application. The current paper provides "design principles" for wavevector dependent spin splitting (SS) of energy bands that parallels the traditional Dresselhaus and Rashba spin-orbit coupling (SOC) -induce splitting, but originates from a fundamentally different source—antiferromagnetism. We identify a few generic AFM prototypes with distinct SS patterns using magnetic symmetry design principles. These tools allow also the identification of specific AFM compounds with SS belonging to different prototypes. A specific compound -- centrosymmetric tetragonal $MnF_2$ -- is used via density functional band structure calculations to quantitatively illustrate one type of AFM SS. Unlike the traditional SOC-induced effects restricted to non-centrosymmetric crystals, we show that antiferromagnetic-induced spin splitting broadens the playing field to include even centrosymmetric compounds, and gives SS comparable in magnitude to the best known ('giant') SOC effects, even without SOC, and consequently does not rely on the often-unstable high atomic number elements required for high SOC. We envision that use of the current design principles to identify an optimal antiferromagnet with spin-split energy bands would be beneficial for efficient spin-charge conversion and spin orbit torque applications without the burden of requiring compounds containing heavy elements.

_________________________________________________________________________
Emails: erashba@physics.harvard.edu ; alex.zunger@colorado.edu

Corresponding authors: Alex Zunger, Zhi Wang

I. Introduction

An electron with momentum $\boldsymbol{p}$ and mass $m$ moving in an inversion symmetry-breaking electric field $\boldsymbol{E}$ in a solid experiences an effective magnetic field $\boldsymbol{B}_{\text{eff}} \sim \boldsymbol{E} \times \boldsymbol{p}/mc^2$ in its rest-frame, where $c$ is the speed of light. In bulk crystals[1] this symmetry breaking electric field is given by the gradient of the crystal potential $\boldsymbol{E} = -\nabla V$, whereas in heterostructures[2] it can be produced by interfacial asymmetry, and in centrosymmetric compounds by the local asymmetry of individual structural sectors[3]. This intrinsic magnetic field couples the electron momentum to its spin, a relativistic effect leading to spin–orbit-coupling (SOC) induced spin splitting of energy bands at wave vectors differing from the time reversal invariant moments (TRIM). In the semi-relativistic Pauli equation, the SOC is described by the Thomas (T)[4] term $H_{\text{T}} = -\frac{e\hbar}{4m^2c^2}[\boldsymbol{\sigma} \cdot (\nabla V(\mathbf{r}) \times \mathbf{p})]$ that couples electron spin $\boldsymbol{\sigma}$ to its coordinate $\boldsymbol{r}$ and momentum $\boldsymbol{p}$, and its fully relativistic generalization. These seminal studies have formed the basis for the development of spintronics[5-7], bringing $\boldsymbol{k}$-dependent spin-orbit interaction to the forefront of solid-state physics, including applications to spin transistor, spin–orbit torque, spin Hall effect, topological insulators, and Majorana Fermions (see review in Ref. [8]).

Since the relativistic SOC increases rapidly with atomic number Z, and since the strength of chemical bonds in compounds decreases rapidly with increasing atomic number (e.g. in the sequence ZnTe-CdTe-HgTe, or ZnS-ZnSe-ZnTe[9]), the ease of breaking such fragile high-Z bonds-- creating vacancies that produce free carriers-- has been an unwelcome but constant companion of high SOC compounds both for spin splitting and for topological insulators applications.[9-12] This double limitation of the Rashba and Dresslhause[1,2] spin splitting effects to high-Z and non-centrosymmetric compounds has limited the playing field, raising hopes for an alternative spin splitting mechanism in thermodynamically stable, low Z compounds of more general symmetries.

More recently, the investigation of spin splitting of energy bands has been expanded to magnetic systems, in particular, antiferromagnets (AFM), for eliminating stray fields.[13-16] For example, spin splitting has been calculated in some high-Z AFM half-metallic compounds such as iron-pnictide AFM $BaCrFeAs_2$[17], $Mn_3Al$ and $Mn_3Ga$[18], and 2D van der Waals[19] AFM materials, but such occurrences were not distinct from the traditional spin orbit effect[1,2]. Indeed, it is generally implied that such splitting in the presence of background AFM may be treated just as SOC-induced splitting in non-magnetic (NM) materials[1,2], through the usual Thomas term[4]. For example, allowing for antiferromagnetism in calculations on $BiCoO_3$[20] having SOC manifests but a small change in its spin splitting; furthermore, if SOC is deliberately removed from the Hamiltonian, the predicted spin splitting vanishes in the whole Brillouin Zone (BZ). Also, the field-free magnetic mechanism discussed in the present paper differs from the anomalous spin-orbit coupling in antiferromagnets induced by applying external magnetic field, discussed in Ref. [21,22].

A phenomenological theory of magnetic spin splitting has been proposed 1964 by Pekar and Rashba [23], suggesting that the presence in magnetic compounds of a spatially-dependent intrinsic magnetic field $\mathbf{h}(\mathbf{r})$, periodic with the crystal period, can lead to coupling of Pauli matrices $\boldsymbol{\sigma}$ to this $\mathbf{h}(\mathbf{r})$. This would result in a magnetic mechanism of $\boldsymbol{k}$-dependent spin splitting, suggestive of a new type of spin orbit coupling. Because the $\boldsymbol{k} \cdot \boldsymbol{p}$ formalism used in Ref. [23] did not afford an atomistic definition of $\mathbf{h}(\mathbf{r})$ and its ensuing spin splitting, nor did it provide for guiding principles to select a target material for investigating such effects, examination of these 1964 ideas remained dormant for a long time.

In the present paper, inspired by Ref. [23], we demonstrate an AFM mechanism that creates $\boldsymbol{k}$-dependent spin splitting $\Delta_{ss}(\boldsymbol{k})$ even in centrosymmetric, low Z compounds, persists even at time reversal invariant wave vectors, and has an unusual quadratic scaling on momentum $\boldsymbol{k}$. The coupling of spin to lattice degrees of freedom via the periodic spatial dependent intrinsic magnetic field $\mathbf{h}(\mathbf{r})$ is analogous to a new form of spin orbit *coupling*; the fact that spin splitting can, however, exist even without the presence of spin orbit *interaction* in the Hamiltonian is noteworthy. We formulate the general magnetic space group conditions ("design principles") for spin splitting in different AFM prototypes, either with or without SOC, and illustrate via detailed first principles calculations a case of purely AFM-induced spin splitting.

## II. Magnetic symmetry conditions for AFM-induced spin splitting

### A. Symmetries that enforce spin degeneracy

To select a compound for direct magnetic $\boldsymbol{k}$-dependent spin splitting we inspect the underlying symmetry requirements. We first list the symmetries that *keep* spin degeneracy, preventing SS, then discuss how to violate those symmetries. (i) As is known[24], the combination $\theta I$ of time reversal $\theta$ and spatial inversion $I$ symmetries ensures double degeneracy for arbitrary wave vector $\boldsymbol{k}$. Likewise, (ii) when SOC is turned off, the spin and spatial degrees of freedom are decoupled, so there could exist pure spin rotation $U$, a spinor symmetry, that reverses the spin but keeps momentum invariance, thus preserving spin degeneracy for all wave vectors. The spin rotation $U$ does not exist in AFM when the alternating magnetic moments reside on different atomic sites, because such arrangement reverses the antiferromagnetic order. But in a specific types of AFM compound (referred to as magnetic space group (MSG) type IV[25], such as BiCoO$_3$[20]) where there exists a translation $T$ that transforms the reversed antiferromagnetic order back, $UT$ symmetry would still preserve spin degeneracy for all wave vectors.

### B. Violating degeneracy-enforcing symmetries

(i) As expected, the appearance of spin splitting requires first the violation of $\theta I$ symmetry. In magnetic crystals, where $\theta$ is already violated due to magnetic order, absence of the inversion $I$ symmetry doesn't mean breaking of $\theta I$, hence does not necessarily lead to the removal of spin degeneracy. (Actually, even for a *centrosymmetric* magnetic structure, where $I$ is preserved but $\theta I$ is broken, one can still have spin splitting). (ii) To have SOC-unrelated spin splitting, one needs also to violate $UT$ symmetry. AFM structures that violate $UT$ symmetry correspond to the so-called MSG type III or I such as rutile MnF$_2$. Appendix A provides more detailed discussion of $UT$ symmetries.

### C. The resulting prototypes of AFM SS

Based on whether the AFM compound in question has or lacks $\theta I$ symmetry, and weather it belongs to MSG type IV or MSG type I / III, we have identified four distinct types of AFM spin splitting prototypes (Table I). The first two prototypes, (1) (2), have spin degeneracy at arbitrary $\boldsymbol{k}$ point because of protection by $\theta I$ symmetry. The prototypes (3) (4) have $\theta I$ violation, allowing spin splitting in the presence of SOC. Prototype (3) being MSG type IV has spin degeneracy when SOC is turned off (referred as "SOC induced spin splitting") whereas prototype (4) being MSG type I or III allows spin splitting even when SOC is turned off (referred as "AFM induced spin splitting"). To find specific compound realizations of the four AFM prototypes (last line of Table I) one can search listings of magnetic symmetries (such as the Bilbao listing[26]) for compliance with our design principles (top 2 lines in Table I). As a concrete example, we illustrate the identification of a realization of AFM SS prototype 4 compound. Tetragonal MnF$_2$ having magnetic space group P4$_2$'/mnm' complies with the above noted design principles -- (1) has no $\theta I$ symmetry despite the presence of inversion symmetry; (2) belongs to MSG type III, therefore no $UT$ symmetry.

**TABLE I | Classification of four spin splitting prototypes in antiferromagnetic compounds in terms of symmetry conditions, consequences, and examples.** *Symmetry conditions:* $\theta$ represents time reversal and $I$ represents spatial inversion, $\theta I$ is the combination of these two operations. AFM can be MSG type I, III or IV. (For detail description of MSG and MSG type please refer to Appendix A). *Consequences:* No SS means no spin splitting either with or without SOC. SOC- induced SS means that one has spin splitting when SOC is non-zero, but no spin splitting when SOC is turned off. AFM induced SS means that one has spin splitting even when SOC is turned off. Note that the symmetry-based conditions generally apply not only to collinear but also to noncollinear AFM. For example, we would expect AFM-induced spin splitting in a non-collinear AFM $Mn_3Ir$[27] which is also centrosymmetric but has no $\theta I$ symmetry and belongs to MSG type III.

| AFM SS prototype | 1 | 2 | 3 | 4 |
|---|---|---|---|---|
| Condition 1: Has $\theta I$? | Yes | Yes | No | No |
| Condition 2: MSG type | III | IV | IV | I or III |
| Consequences | No SS | No SS | SOC induced SS | AFM induced SS |
| Examples | CuMnAs[28] | NiO[29] | $BiCoO_3$[30] | $MnF_2$ |

Table I indicates that not all AFM compounds have the same SS behavior, and that the magnetic, not just spatial symmetries are important. For example, an AFM SS has been theoretically analyzed recently based on tight-binding models on the multipole description by Hayami et. al.[31,32]. However, their multipole analysis was based on point group symmetry not magnetic group symmetry, omitted the non-magnetic atoms. This omission (e.g. $MnF_2$ without F), however, restores the $UT$ symmetry and results in the prediction of complete spin degeneracy in the absence of SOC, in sharp contrast with DFT predicted (below) giant spin splitting. (see Appendix B for detail discussion of these previous work)

## III. Illustration of the properties of an AFM-induced SS compound MnF2

### A The system

$MnF_2$ is a wide gap insulator both below and above its Néel temperature of 67K.[33] It is a centrosymmetric rutile structure (conventional space group $P4_2/mnm$), with magnetic Mn ions occupying position (0, 0, 0) and (1/2, 1/2, 1/2) centered in an octahedral of non-magnetic F anions located at ±(u, u, 0) and ±(1/2+u, 1/2-u, 1/2) where u is the positional parameter. The refinement X-ray diffraction results[34] gave the positional parameter u=0.305, and lattice constants a=b=4.873 Å, c=3.311 Å. Erickson[35] found via neutron scattering measurements the AFM moment aligned along the tetragonal axis (i.e., [001]) with magnetic space group of $P4_2'/mnm'$. The magnetic crystal unit cell is shown in Figure 1(a). While concentrating on this specific material $MnF_2$ as an illustration of a new physical effect, we do not maintain that it is optimized for technological usage in a specific spintronics device application (size of band gap; dopability; value of Néel temperature). Optimization of such material constants might be possible by comparing different compounds belonging to a given AFM SS prototype. This is outside the scope of the current paper.

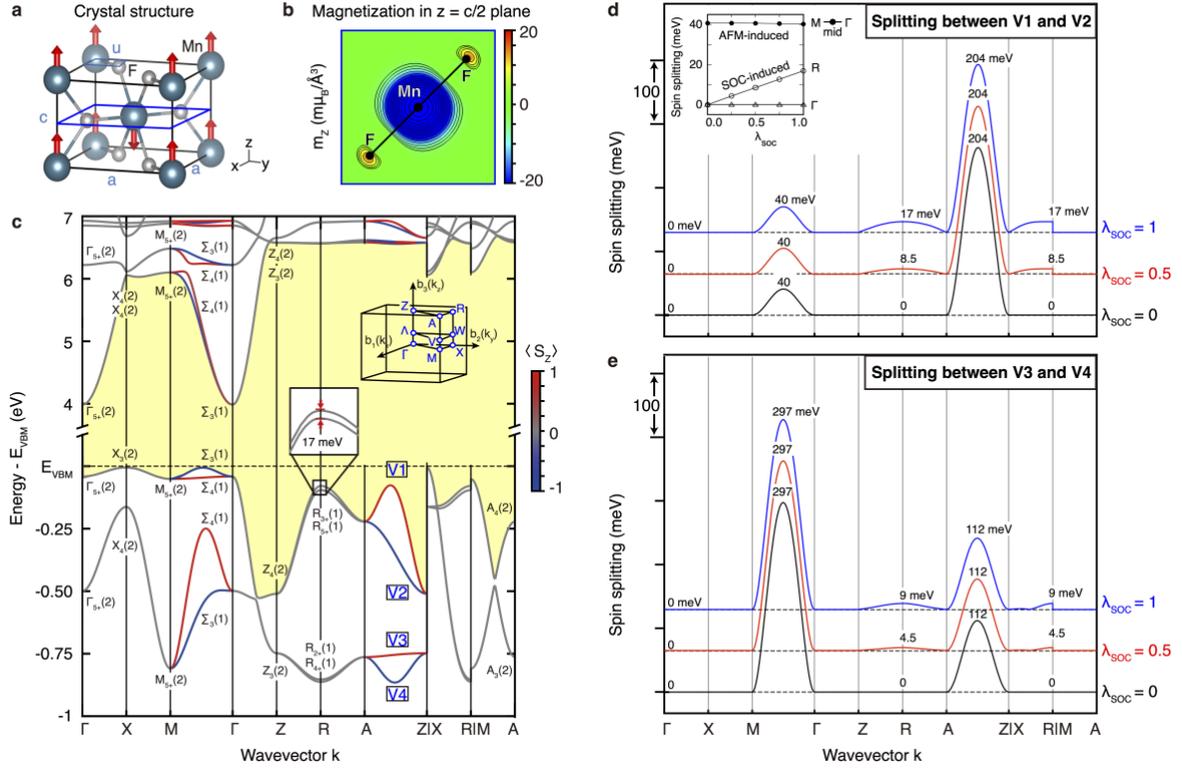

**Figure 1 | Crystal structure, band structure and spin splitting of the centrosymmetric AFM tetragonal MnF$_2$.** **(a)** Magnetic unit cell where red arrows indicate local magnetic moment; **(b)** contour plot of magnetization along z in $z = c/2$ plane; **(c)** DFT calculated band structure with our calculated magnetic symmetry representations (see Appendix C), using the notations of Ref. [36] with numbers in parenthesis indicating the dimension of the representation (i.e., degeneracies). The top four valence bands are denoted by V1, V2, V3, V4 and the yellow screen highlights the gaps between valence and conduction bands. Insert of (c) shows the BZ and the blow-up bands around R point. The blue to red color scale denotes the calculated out-of-plane spin polarization. Panels **(d, e)** show DFT calculated wave vector dependence of the spin splitting between pairs of valence bands V1-V2 (in (d)) and between V3-V4 (in (e)) for different scaling of SOC $\lambda_{SOC}$ (numerical coefficient $0 < \lambda_{SOC} < 1$). Insert of (d) shows the spin splitting vs. the amplitude of the spin orbit coupling $\lambda_{SOC}$ at Γ (0, 0, 0), R (0, 0.5, 0.5) and the middle point of Γ-M (0.25, 0.25, 0). All DFT calculations use PBE exchange correlation functional[37] with on-site coulomb interaction on Mn-3d orbitals of U= 5eV, J= 0 eV and the experimental crystal structure[34].

## B    Six predicted characteristics of AFM-induced spin splitting in MnF$_2$

We calculated the relativistic electronic structure of AFM MnF$_2$ within density functional theory (DFT) (see description of DFT method in Appendix C). Figure 1(b) provides the calculated magnetization $m_z(r) = m^\uparrow(r) - m^\downarrow(r)$ in the z=c/2 plane, with $m^\uparrow$ and $m^\downarrow(r)$ representing the up and down spin electron density. To assess the AFM magnetism effect, we also define a reference NM model, where the magnetic moment on each site is zero, resulting in a metallic state. We emphasize that the NM model is not used to mimic the physical high temperature paramagnetic (PM) phase that has a distribution of non-vanishing local magnetic moments that creates an insulating gap even in the absence of long-range order.[38,39] Figure 1(c) gives the band structure of the AFM phase calculated with SOC in its experimental crystal structure. We find a z-oriented magnetic moment on $Mn^{2+}$ of 4.7 $\mu_B$, in good agreement with the neutron scattering measurement of 4.6 $\mu_B$. We also find calculated minimum direct gap at Γ of 4.02 eV and a smaller indirect gap between VBM at X and CBM at Γ of 3.98 eV, comparable with the measured absorption gap[40] of 4.1 eV (estimated from the convergence limit of the observed series of discrete d-d* multiplet transitions into the onset of band-to-band continuum). The DFT (mean-field) calculated band gap and DFT local moment both agree with experiment, providing strong evidence that the single-particle band structure picture with a 5 eV wide band width as advanced in the DFT calculations holds well, supporting the notion of well-defined coherent bands.

To assist future measurements of the predicted AFM-induced ***k***-dependent spin splitting (e.g. via angle-resolved photoemission spectroscopy (ARPES) and spin-ARPES) as well as potential applications in novel spintronics we next describe the main predicted features of the AFM-induced spin splitting:

*(i) The spin splitting has a typical atomic-like energy scale ("giant splitting"):* Despite rather small atomic numbers in $MnF_2$ (Z(Mn)=25 and Z(F)=9), the magnitude of the spin splitting (up to 300 meV seen between V3 and V4 along Γ-M in Figure 1(e)) arising from the AFM mechanism can be comparable to some of the largest known spin splitting of conventional electric mechanism for heavy atom high Z compounds, such as the 'giant SOC' induced spin splitting in BiTeI[41] and GeTe[42,43]. The reason for the difference is that the magnetic field which induces the splitting in AFM reflects the local magnetic moments *localized about atomic sites*, not as in the SOC effect where the inducing magnetic field reflects the asymmetry in the *inter-atomic regions* of the unit cell. The locality of magnetic moments needed for obtaining large spin splitting does not contradict the requirement to introduce itinerant carriers. Local magnetic moments of 4-5 $\mu B$ are common in Mn-salts with broad (4-5 eV) bands and high electronic mobility, e.g., $La_{1-x}Sr_xMnO_3$ [44]. We find that the spin split bands in $MnF_2$ occur about 40 meV below the VBM (bands V1-V2 in Figure 1) and about 500 meV below the VBM (bands V3-V4 in Figure 1). Either should be amenable to photoemission detection for validating the theory.

*(ii) The splitting persists even if SOC= 0*: The spin splitting along the Γ-M and Z-A lines is present even when SOC is turned off in the Hamiltonian (black line in Figure 1(d) and (e); also shown in the insert of (d)). This is very different from prototype 3 AFM (Table I) $BiCoO_3$[20], where spin splitting disappear**s** if SOC vanishes. Thus, the AFM-induced spin splitting mechanism delivers the long-standing hope for wave vector dependent spin splitting mechanism in thermodynamically stable, low Z compounds.

*(iii) Relative to the NM case, AFM induces a highly anisotropic and **k**-dependent spin splitting*: We show in Figure 2 the band structures of centrosymmetric $MnF_2$ in two cases: (a) NM without SOC; (b) AFM without SOC. In both cases we indicate the degeneracies of states, calculated by DFT shown as integer values. An important manifestation of the AFM-induced spin splitting (Figure 1(c)) is that whereas in the NM structure, the whole BZ, including directions Γ-X and Γ-M, have doubly degenerate (non-split) bands, in the AFM structure spin splitting arises even in the absence of SOC but it is wave vector dependent. Bands remain degenerate along the Γ-X directions, but become spin split along the Γ-M direction. Such anisotropic spin splitting was already hinted by the asymmetry in magnetization in coordinate space as shown in Figure 1(b) between $\boldsymbol{x+y}$, $\boldsymbol{x-y}$ and $\boldsymbol{x}, \boldsymbol{y}$ directions. This behavior is understandable on the basis of magnetic symmetry (See Appendix D for discussion of unitary and antiunitary symmetries): the AFM ordering does not lead to symmetry reduction along the Γ-X paths, relative to its NM counterpart. The resulting spin degeneracy along $k_x$ (or $k_y$) direction of Γ-X in AFM is protected by its group of ***k*** symmetries $\theta\{C_{2x}|\boldsymbol{\tau}\}$ and $\theta\{\sigma_{vy}|\boldsymbol{\tau}\}$ (or $\theta\{C_{2y}|\boldsymbol{\tau}\}$ and $\theta\{\sigma_{vx}|\boldsymbol{\tau}\}$). In contrast, along the Γ-M paths, in AFM the combined symmetries of $\theta\{C_{2a}|0\}$ and $\theta\{C_{2b}|0\}$ (or $\theta\{\sigma_{da}|0\}$ and $\theta\{\sigma_{db}|0\}$) are broken, which creates spin splitting. Here, $C_{2x}, C_{2y}, C_{2a}, C_{2b}$ are π rotations about the [100], [010], [110], [1-10] axes,

respectively; $\sigma_{vx}, \sigma_{vy}, \sigma_{da}, \sigma_{db}$ are mirror reflections in (100), (010), (1-10), (110) planes; and vector $\tau$ = (1/2,1/2,1/2) is half lattice translation, directed along the spatial diagonal [111] of the unit cell. Similar arguments (given in Appendix D) apply for spin degeneracy along Z-R and spin splitting along Z-A in Figure 1(b).

*(iv) The AFM mechanism gives rise to even powers of k in the spin splitting:* Of special interest in Figure 1 (c) is the diagonal Γ-M and Z-A lines showing large spin splitting while at the end of these $k$-lines the splitting vanishes. It is of interest therefore, to establish how the splitting changes near its $k$-space end points. By fitting the DFT calculated spin splitting $\Delta_{ss}(k)$ to the power of k, we found a quadratic-in-k dependence at $\Gamma_{\Gamma-M}$ (i.e., near the Γ-end along $\Gamma - M$ path) and $M_{M-\Gamma}$ (see details of fitting in Appendix E and effective model in Appendix F). Thus, near degeneracy points $\Delta_{ss}(k)$ has a quadratic k dependence compared with odd powers typical of the electrically induced SOC effect.

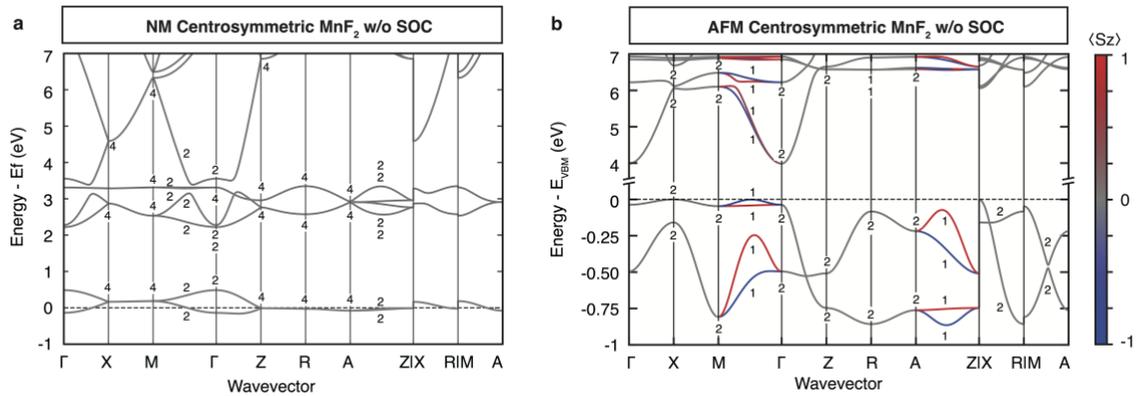

**Figure 2 | DFT band structures of centrosymmetric (CS) MnF$_2$ in NM and AFM without SOC.** In all cases we use the experimentally observed centrosymmetric tetragonal structure[34]: **(a)** NM with SOC set to zero; **(b)** AFM with SOC set to zero. Out-of-plane spin polarizations are mapped to color scale from blue to red. The integer number attached to each band and $k$ point is the degeneracies.

*(v) A Dresselhaus in-plane spin texture results from a cooperative SOC and AFM effect:* The coupling between spin space and position space results not only in spin-splitting of the energy spectrum, but also in developing "spin-momentum locking", where the spin orientation is locked with momentum $k$. The vector field of the spin states in momentum space is called spin texture, being helical for the conventional Rashba SOC mechanism[2] and non-helical for the Dresselhaus mechanism[1]. The spin texture for AFM-induced spin splitting has its own fingerprints. Figure 3 shows the calculated spin textures of the V1 and V2 bands at the representative $k$-plane $k_z = \pi/2c$ where $c$ is the lattice constant along $z$ axis. We see that, electron spins are mostly aligned along the out-of-plane z direction, as can be surmised from the magnetic structure (see Figure 1(a)). This is seen in the four quadrants patterns on a fixed $k_z$ plane with positive (up arrow in Figure 3(a) and (b)) and negative (down arrow in Figure 3(a) and (b)) out-of-plane spin polarizations in the neighbor quadrants. The out-of-plane spin polarizations are opposite in sign between bands V1 and V2, as noted by the reversal of the red and blue patterns for V1 and V2. Similar four quadrants pattern of out-of-plane spin polarization is also found in the $k_z = 0$ and $k_z = \pi/c$ planes (see corresponding spin texture results in Appendix G).

Interestingly, inspecting the $k_z = \pi/2c$ plane, Figure 3 shows a pronounced (i) in-plane (ii) non-helical Dresselhaus-like spin texture. These features are unexpected given that the crystal structure of MnF$_2$ is magnetized in the z-direction and centrosymmetric, while normally to assure Dresselhaus features[3] we need non-centrosymmetric symmetry. We find that the Dresselhaus spin texture in MnF2 requires for its existence the SOC term (i.e. the texture vanishes if the SOC is removed from the Hamiltonian). Thus, the texture represents the combined effect of coexistence of SOC with AFM (see cooperative effect on spin splitting in Appendix H).

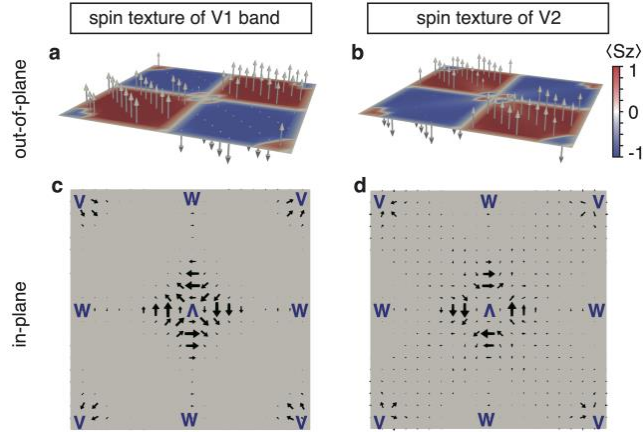

**Figure 3 | Spin textures in AFM MnF$_2$ with SOC on $k_z = \pi/2c$ plane**. **(a)** Out-of-plane spin texture of V1 band, **(b)** out-of-plane spin texture of V2 band, **(c)** in-plane spin texture of V1 band, and **(d)** in-plane spin texture of V2 band.

*(vi) Different wavevectors can have different dependence on SOC strength:* The insert of Figure 1(d) shows different characteristic behaviors of the dependence of spin splitting $\Delta_{ss}(\boldsymbol{k})$ on spin-orbit strength at different $\boldsymbol{k}$ points: (1) The trivial case (e.g. Γ point) is that neither magnetic nor SOC induces any splitting; (2) the R point shows zero spin splitting when $\lambda_{SOC} = 0$ and linear dependence of $\lambda_{SOC}$, illustrating a cooperation of both magnetic and SOC mechanism; notice that despite R being a TRIM point, it shows spin splitting, unlike the case of purely SOC induced effects; (3) the non-trivial case of purely magnetic induced spin splitting occurs along Γ-M (as well as A-Z) line, where non-zero spin splitting is present even at $\lambda_{SOC} = 0$ and is almost independent of $\lambda_{SOC}$. The appearance of such distinct spin splitting behaviors at different wave vectors in a single compound would be advocated for multifunctional spintronic applications.

## IV.     Discussion

This study uncovers the design principles of spin splitting in AFM compounds based on magnetic symmetry analysis and shows a very rich set of fingerprint fundamental physical effects ((i)-(vi) above) in a specific prototype, including the giant spin splitting that characterizes the AFM mechanism and could aid its future experimental observation. The present symmetry-based theory with atomistic resolution enabled by DFT instills content into the 1964 phenomenological theory by Pekar and Rashba[23] proposing a pioneering magnetic spin splitting mechanism.

The mechanism described here foresees many encouraging physical phenomena. As an example, active research is going currently on 2D layered systems consisting of two layers, of which one is an AFM and the other a heavy metal such as Pt, with the SOC of Rashba-type developing on their interface and controlled by electric bias applied across it.[45] [46] Using antiferromagnets with spin-split bands, which in addition are either magneto-electric or piezoelectric, might eliminate necessity of the heavy-metal layer due to the giant magnitude of spin-orbit splitting found above. Similarly, we expect the mechanism accounting for the giant anisotropy of gilbert damping predicted in AFM[47]. We also note a few transport effects that are likely associated to the AFM-induced spin splitting effect. These include finite spin current predicted by Yan et. al.[48], anomalous Hall conductivity predicted by MacDonald et. al.[49] and Arita et. al.[50], and crystal Hall effect proposed[51] and verified[52] by Jungwirth and Sinova et. al.

**Acknowledgments**

The National Science Foundation (NSF) Grant NSF-DMR-CMMT No DMR-1724791 supported the theory development of this work by L.-D.Y., Z.W., and A.Z. at the University of Colorado Boulder. The *ab initio* calculations of this work were supported by the U.S. Department of Energy, Office of Science, Basic Energy Sciences, Materials Sciences and Engineering Division under Grant No. DE-SC0010467. J.-W. L. was supported by the National Natural Science Foundation of China (NSFC) under Grant Number 61888102. This work used resources of the National Energy Research Scientific Computing Center, which is supported by the Office of Science of the U.S. Department of Energy under Contract No. DE-AC02-05CH11231. We thank Dr. Carlos Mera Acosta for fruitful discussions.

**APPENDIX A: Additional symmetry considerations beyond $\theta I$ for selecting compounds having non-zero magnetic-induced spin splitting**

To select magnetic crystals that have non-zero contribution from the magnetic mechanism means to find the crystals that still have ***k***-dependent spin splitting even in the absence of SOC (i.e., when there is no contribution from the electric mechanism). To do so one must violate $UT$ symmetry, where $U$ is a spinor symmetry of **SU(2)** which reverses the spin state and $T$ is a translation of the primitive lattice. This requirement stems from the fact that the existence of $UT$ symmetry preserves the double degeneracy in the whole Brillouin zone, as it transfers any spin state to its opposite spin state while keeping ***k***-invariant for arbitrary wavevector. Such $UT$ symmetry exists in AFM compounds whose magnetic unit cell is not equivalent to its non-magnetic unit cell: the primitive lattice translation $T$ translate those up-spin (down-spin) atoms to occupy the down-spin (up-spin) atom sites while $U$ reverses the spin, thus $UT$ symmetry preserves the crystal structure.

Antiferromagnets with primitive lattice translations that reverse the microscopic magnetic moments are formally known as having black-and-white Bravais lattice and type IV in terms of magnetic space group (MSG). Formally, the MSG includes not only the unitary symmetries (US), i.e., spatial symmetries, but also antiunitary symmetries (AS), that are time reversal $\theta$ and its combination with spatial operations.

In terms of its construction of the unitary and antiunitary part from the space group (G), MSG can be classified into four types.

- MSG that has no antiunitary symmetries (i.e., AS = ∅) are identified as MSG type I.

- MSG that has the unitary part equivalent to G and an equal number of antiunitary symmetries, *i.e.,* US = G, AS = $\theta$G belongs to MSG type II. Since all NM cases have time reversal symmetry, they all belong to this category;
- MSG that has the unitary part composed of half of its space group symmetries (spatial operations that keep the atomic structure invariant) are MSG type III or type IV: If the system has a normal Bravais lattice (magnetic unit cell equivalent to its NM primitive unit cell) it is MSG type III; if the system has a black and white Bravais lattice (magnetic unit cell being supercell of its NM primitive unit cell) it is MSG type IV.

AFM with antiferromagnetic order spontaneously breaks time reversal symmetry, therefore can't be MSG type II but can be MSG type I, III or IV.

Our selected AFM compound $MnF_2$ has magnetic space group of $P4_2'/mn'm$ without SOC and the space group $P4_2'/mnm'$ with SOC, both of which belong to MSG type III with equivalent AFM unit cell ($Mn_2F_4$) to the NM primitive unit cell ($Mn_2F_4$). Therefore, it is expected to have spin splitting in such AFM compound from symmetry perspective.

**APPENDIX B: Previous studies on spin splitting in AFM compounds**

There are many previous works on spin splitting in AFM compounds. While most of the studies mentioned the occurrence of spin splitting, only a few literatures tried to establish a causal understanding of such phenomenon. To give a clear view of the conceptional advance of this paper, here we list several previous studies on the spin splitting in AFM and compare them to this work.

**(a) Previous studies on the occurrence of spin splitting in AFM compounds.**

Hu *et al.* [17] predicted that iron-pnictide AFM $BaCrFeAs_2$ could be half-metallic due to the spin splitting, by first-principles and tight-binding calculations. Gao *et al.* [18] predicted AFM $Mn_3Al$ and $Mn_3Ga$ to have spin splitting by first-principles calculation. Gong *et al.* [19] showed that 2D van der Waals AFM materials could have spin splitting under density functional theory. In this work, we are not satisfied with only showing the existence of spin splitting in AFM without the need of SOC (to aid experimental testing); more importantly, we formulate the fundamental magnetic space group conditions ("design principles") for spin splitting in the absence of SOC – violation of both $\theta I$ and $UT$ symmetries.

**(b) Previous studies on the causal understanding of spin splitting in AFM.**

As far as we know, the first literature of the causal understanding was from one of our coworker Emmanuel Rashba[23], which has already been discussed in the main text. Other previous works include: Hayami *et al.*[31] examined AFM spin splitting when SOC is absent, by using tight-binding model on the multipole description based on point group symmetry of the magnetic element. In this work we have found that the complete symmetry analysis should also include non-magnetic atoms and the translational symmetry, meaning the magnetic group symmetry. For example, in the case of tetragonal $MnF_2$, removing the non-magnetic F- anions will restore the $UT$ symmetry and result in a complete spin degeneracy in the absence of SOC, in sharp contrast our DFT predicted giant spin splitting. To address the real-life case where SOC is finite, in this work we also have discussed both cases when SOC is present and absent.

Another previous work by Naka et al. [32] shows in a specific type of organic antiferromagnet the spin splitting effect can be used for spin current generation. The authors described the spin splitting effect as originating from unspecified AFM order induced by real space molecular arrangement anisotropy in a class of organic antiferromagnets. However, the descriptive understanding is specific to organic antiferromagnet with checker-plate-type lattice. While in this work we have offered more general design

## APPENDIX C: DFT Calculation methods and parameters

**3.1 DFT calculation parameters:** We have studied the electronic and spin properties of MnF$_2$ using DFT with the experimental crystal structure[34] and the experimental spin configuration[35]. In the DFT calculations, we use Perdew-Burke-Ernzerhof (PBE) exchange correlation functional[37] with plane wave basis (energy cutoff of 500 eV and $10 \times 10 \times 14$ Monkhorst-pack $k$-mesh sampling[53]). We use the on-site potentials of U=5 eV and J=0 eV on Mn 3d orbitals following Liechtenstein approach[54].

**3.2 Controlling SOC strength in AFM and NM:** Band structures under different (artificial) SO strengths are calculated by introducing a numerical pre-factor $\lambda_{SOC}$ ($0 < \lambda_{SOC} < 1$) to the SO Hamiltonian term $\lambda_{SOC} \left( \frac{\hbar}{2m_e^2 c^2} \frac{K(r)}{r} \frac{dV(r)}{dr} \hat{L} \cdot \hat{S} \right)$ in the DFT formalism[55], where $\hat{L} = \hat{r} \times \hat{p}$ is the orbital angular momentum operator, $\hat{S}$ is the spin operator, $V(r)$ is the spherical part of the effective all-electron potential within the projector augmented plane wave (PAW) sphere, and $K(r) = \left(1 - \frac{V(r)}{2m_e c^2}\right)^{-2}$. Hamiltonian (as well as wavefunctions, charge density, and $V(r)$, *etc.*) are still calculated self-consistently. To study the effect of magnetic mechanism we use a reference NM model where the magnetic moment on each atomic site is zero. We emphasize that the NM model is not used to mimic the physical high temperature paramagnetic phase that has zero total moment but a distribution of non-vanishing local magnetic moments that creates an insulating gap even in the absence of long-range order.[38,39]

**3.3 Spin polarization and spin texture calculations:** The spin polarization for Bloch state $|k\rangle$ is calculated via the definition of $\langle k|\hat{S}|k\rangle$, which can be decomposed into two components: the out-of-plane spin polarization $\langle k|\hat{S}_z|k\rangle$ and the in-plane spin polarization $\langle k|(\hat{S}_x, \hat{S}_y)|k\rangle$. Spin texture of selected band on $k$-plane is calculated by evaluating both the out-of-plane and in-plane spin polarizations for each $|k\rangle$ on the $k$-plane.

**3.4 Representations for bands:** The double group irreducible (co) representation for each degenerate state at the high symmetry $k$ point of the AFM phase with SOC is derived by: (a) we first calculate the transformation properties of selected Bloch basis states under the relevant group of $k$ symmetries; the Bloch basis is constructed to have the same spin and orbital character (obtained from DFT) and the ensuring transformation properties of the degenerate states; (b) we then identify and label the degenerate bands at such $k$ points adopting the names of irreducible (co) representations for MnF$_2$ from Ref. [36]; (c) whether additional degeneracy will be induced by antiunitary symmetries is determined using Wigner's test[25] given in Ref. [36].

## APPENDIX D: Symmetry analysis of band eigenstates in MnF$_2$

The spin degeneracy and splitting are direct consequences of symmetry preservations and reductions upon introducing AFM and SOC. We see that it is the introducing of antiferromagnetic order from NM to AFM phase of MnF$_2$ that breaks the four-fold axial symmetry and makes directions <100> and <110> non-equivalent. Such symmetry breaking manifests itself dramatically in the anisotropic spin splitting of electron bands; see in Figure 1 (c).

**4.1 Symmetry protected spin degeneracy:** Given the Hamiltonian $\widehat{H}$ and one of its eigenvectors $\psi$ with eigenvalue $E$, for any symmetry $\hat{g}$ of $\widehat{H}$ (that has $[\hat{g},\widehat{H}] = 0$), $\hat{g}\psi$ is also an eigenvector of $\widehat{H}$ with the same eigenvalue $E$. This is easily verified as:

$$\widehat{H}\hat{g}\psi = \hat{g}\widehat{H}\psi = E\hat{g}\psi \quad (S1)$$

When $\psi$ and $\hat{g}\psi$ are linear independent states, they form a pair of degenerate states; the spin degeneracy at specific $\boldsymbol{k}$ points can then be protected if $\hat{g}$ also keeps $\boldsymbol{k}$ invariant, i.e., $\hat{g}\boldsymbol{k} = \boldsymbol{k} + \boldsymbol{G}$ (where $\boldsymbol{G}$ is the reciprocal lattice vector). For example, for $\hat{g}$ being the TR symmetry and $[\hat{g},\widehat{H}] = 0$, $\hat{g}$ enforces doubly spin degeneracy at TRIM points.

**4.2 Space groups and symmetry operators for MnF$_2$ NM and AFM phases, with and without SOC:** If one does not consider the time reversal symmetry $\theta$, the space group $G$ of NM MnF$_2$ is P4$_2$/mnm, consisting of 16 unitary symmetries. Using the subgroup $H = \{E, C_2, I, \sigma_h\}$ of $G$ (index $[G:H] = 4$), we can write the partition of $G$ using $H$ and its three left cosets $LH1$, $LH2$, and $LH3$ as listed in Table SII:

**Table SI | Explicit lists of space group symmetries of NM MnF$_2$.**

| $H$ | $LH1 = \{C_{2a}\|0\}H$ |
|---|---|
| $\{E\|0\}: (x,y,z) \to (x,y,z)$ | $\{C_{2a}\|0\}: (x,y,z) \to (y,x,-z)$ |
| $\{C_2\|0\}: (x,y,z) \to (-x,-y,z)$ | $\{C_{2b}\|0\}: (x,y,z) \to (-y,-x,-z)$ |
| $\{I\|0\}: (x,y,z) \to (-x,-y,-z)$ | $\{\sigma_{da}\|0\}: (x,y,z) \to (y,x,z)$ |
| $\{\sigma_h\|0\}: (x,y,z) \to (x,y,-z)$ | $\{\sigma_{db}\|0\}: (x,y,z) \to (-y,-x,z)$ |
| $LH2 = \{C_{2x}\|\boldsymbol{\tau}\}H$ | $LH3 = \{C_4\|\boldsymbol{\tau}\}H$ |
| $\{C_{2x}\|\boldsymbol{\tau}\}: (x,y,z) \to \left(x+\frac{1}{2},-y+\frac{1}{2},-z+\frac{1}{2}\right)$ | $\{C_4\|\boldsymbol{\tau}\}: (x,y,z) \to \left(-y+\frac{1}{2},x+\frac{1}{2},z+\frac{1}{2}\right)$ |
| $\{C_{2y}\|\boldsymbol{\tau}\}: (x,y,z) \to \left(-x+\frac{1}{2},y+\frac{1}{2},-z+\frac{1}{2}\right)$ | $\{C_4^-\|\boldsymbol{\tau}\}: (x,y,z) \to \left(y+\frac{1}{2},-x+\frac{1}{2},z+\frac{1}{2}\right)$ |
| $\{\sigma_{vx}\|\boldsymbol{\tau}\}: (x,y,z) \to \left(-x+\frac{1}{2},y+\frac{1}{2},z+\frac{1}{2}\right)$ | $\{S_4\|\boldsymbol{\tau}\}: (x,y,z) \to \left(y+\frac{1}{2},-x+\frac{1}{2},-z+\frac{1}{2}\right)$ |
| $\{\sigma_{vy}\|\boldsymbol{\tau}\}: (x,y,z) \to \left(x+\frac{1}{2},-y+\frac{1}{2},z+\frac{1}{2}\right)$ | $\{S_4^-\|\boldsymbol{\tau}\}: (x,y,z) \to \left(-y+\frac{1}{2},x+\frac{1}{2},-z+\frac{1}{2}\right)$ |

Here $E$ is the identity; $I$ is the spatial inversion; $C_2, C_{2x}, C_{2y}, C_{2a}, C_{2b}$ are $\pi$ rotations about the [001], [100], [010], [110], [1-10] axes, respectively; $\sigma_h, \sigma_{vx}, \sigma_{vy}, \sigma_{da}, \sigma_{db}$ are reflections in (001), (100), (010), (1-10), (110) planes, respectively; $C_4$ and $C_4^-$ are counterclockwise and clockwise $\pi/2$ rotations about the (001) axis; $S_4$ and $S_4^-$ are counterclockwise and clockwise $\pi/2$ rotations about the (001) axis followed by an inversion; vector $\boldsymbol{\tau}$ = (1/2,1/2,1/2) is half lattice translation, directed along the spatial diagonal [111] of the unit cell.

When considering the time reversal symmetry $\theta$, the 16 unitary symmetries form an subgroup $G_U = H + LH1 + LH2 + LH3$, while the NM system also has $\theta$ combined with all 16 unitary symmetries in $G_U$, leading to an anti-unitary set $G_{AU}$. Using the prime symbol to indicate time reversal symmetry, we have $G_{AU} = G_U' = \theta G_U$. The entire group now becomes $G = G_U + G_{AU} = G_U + G_U' = H + LH1 + LH2 + LH3 + H' + LH1' + LH2' + LH3'$

Including SOC in NM couples the spatial rotation to spin rotation, which results in a double space group of P4$_2$/mnm composed of $H, LH1, LH2, LH3, H', LH1', LH2', LH3'$ and their combination with a rotation of $2\pi$ ($\bar{E}$): $H_D = \{\bar{E}, \overline{C_2}, \bar{I}, \overline{\sigma_h}\}$, $LH1_D = \{C_{2a}|0\}H_D$, $LH2_D = \{C_{2x}|\boldsymbol{\tau}\}H_D$, $LH3_D = \{C_4|\boldsymbol{\tau}\}H_D$, and $H'_D$, $LH1'_D$, $LH2'_D$, $LH3'_D$. In centrosymmetric MnF2, due to the presence of $\theta I$ symmetry ($\theta I \in H'_D$), all bands are spin degenerates within the whole BZ.

Going from NM to AFM, the above P42/mnm space group reduces to a magnetic group $M$ consisting of unitary $G_U$ and antiunitary $G_{AU}$ parts $M = G_U + G_{AU}$. In the absence of SOC in AFM, the unitary part $G_U = H + LH1$, while the antiunitary part is $G_{AU} = LH2' + LH3'$. Including SOC in AFM couples the spatial rotation to spin rotation in the manner of one to two mapping from **SO(3)** to **SU(2)**; thus, the rotations of $LH1$ and $LH2$ not only rotate the spatial space but also reverse the spin orientation. Consequently, the unitary part of the magnetic space group becomes $G_U = H + LH2 + H_D + LH2_D$, and the antiunitary part becomes $G_{AU} = LH1' + LH3' + LH1'_D + LH3'_D$. The above symmetry analysis of MnF$_2$ is summarized in Table SIII.

**Table SII | Unitary and anit-uniary symmetries of MnF$_2$ with inclusion and exclusion of SOC in NM and AFM phases.**

| MnF$_2$ | w/o SOC | w/ SOC |
| --- | --- | --- |
| NM | **Space group:** P4$_2$/mnm<br>$G_U$: $H, LH1, LH2, LH3$<br>$G_{AU}$: $H', LH1', LH2', LH3'$ | **Double space group:** P4$_2$/mnm<br>$G_U$: $H, LH1, LH2, LH3, H_D, LH1_D, LH2_D, LH3_D$<br>$G_{AU}$: $H', LH1', LH2', LH3', H'_D, LH1'_D, LH2'_D, LH3'_D$ |
| AFM | **Magnetic space group:** P4$_2$'/mn'm<br>$G_U$: $H, LH1$<br>$G_{AU}$: $LH2', LH3'$ | **Magnetic double space group:** P4$_2$'/mnm'<br>$G_U$: $H, LH2, H_D, LH2_D$<br>$G_{AU}$: $LH1', LH3', LH1'_D, LH3'_D$ |

**4.3 Different spin splitting behaviors along Γ-X, Γ-M, and Z-R directions in AFM MnF$_2$:** Along the spin degenerate $k$-path Γ-X, the coordinate of $\boldsymbol{k}$ is $(u, 0, 0)$ with $u$ an arbitrary real value between 0 and 1/2. The possible symmetries that keep $\boldsymbol{k}$ invariant are (notice that all conclusions below also are applicable for $\boldsymbol{k} = (0, u, 0)$ by interchanging $x$ with $y$):

$$\{E|0\}: (u, 0, 0) \to (u, 0, 0)$$
$$\{\sigma_h|0\}: (u, 0, 0) \to (u, 0, 0)$$
$$\{C_{2x}|\boldsymbol{\tau}\}: (u, 0, 0) \to (u, 0, 0) \qquad (S2)$$
$$\{\sigma_{vy}|\boldsymbol{\tau}\}: (u, 0, 0) \to (u, 0, 0)$$
$$\theta\{C_{2y}|\boldsymbol{\tau}\}: (u, 0, 0) \to (u, 0, 0)$$
$$\theta\{\sigma_{vx}|\boldsymbol{\tau}\}: (u, 0, 0) \to (u, 0, 0)$$

(1) When SOC is ignored, the magnetic space group is P4$_2$'/mn'm, among above 6 symmetries only $\{E|0\}, \{\sigma_h|0\}, \theta\{C_{2y}|\boldsymbol{\tau}\}$, and $\theta\{\sigma_{vx}|\boldsymbol{\tau}\}$ are symmetries of the magnetic system, where both $\theta\{C_{2y}|\boldsymbol{\tau}\}$ and $\theta\{\sigma_{vx}|\boldsymbol{\tau}\}$ will transfer spin state to opposite spin state and enforce degeneracy between them.

(2) When SOC is considered, the magnetic space group is P4$_2$'/mnm', among above 6 symmetries only unitary symmetries $\{E|0\}, \{\sigma_h|0\}, \{C_{2x}|\boldsymbol{\tau}\}$, and $\{\sigma_{vy}|\boldsymbol{\tau}\}$ are symmetries of the magnetic system, where either $\{C_{2x}|\boldsymbol{\tau}\}$ or $\{\sigma_{vy}|\boldsymbol{\tau}\}$ will transfer the spin eigenstate to a linearly independent spin state, therefore enforcing spin degeneracy between them. The same conclusion can also be obtained from the fact that the group of the wavevector formed by the four unitary symmetries has only one 2D double group irreducible representation $\Delta_5$ (see Table III in Ref. [36]).

Along the spin splitting k-path Γ-M, the coordinate of $\bm{k}$ is $(u, u, 0)$. The possible symmetries that keep $\bm{k}$ invariant are (notice that all conclusions below also are applicable for $\bm{k} = (-u, u, 0)$ by interchanging *a* with *b*):

$$\begin{aligned}
\{E|0\}: (u, u, 0) &\rightarrow (u, u, 0) \\
\{\sigma_h|0\}: (u, u, 0) &\rightarrow (u, u, 0) \\
\{C_{2a}|0\}: (u, u, 0) &\rightarrow (u, u, 0) \\
\{\sigma_{db}|0\}: (u, u, 0) &\rightarrow (u, u, 0) \\
\theta\{C_{2b}|0\}: (u, u, 0) &\rightarrow (u, u, 0) \\
\theta\{\sigma_{da}|0\}: (u, u, 0) &\rightarrow (u, u, 0)
\end{aligned} \quad (S3)$$

(1) When SOC is ignored, the group of the wavevector only has four unitary symmetries $\{E|0\}$, $\{\sigma_h|0\}$, $\{C_{2a}|0\}$, and $\{\sigma_{db}|0\}$, none of these would reverse the spin state and therefore spin splitting is expected along this direction.

(2) When SOC is considered, the group of the wavevector has two unitary symmetries $\{E|0\}$, $\{\sigma_h|0\}$, and two antiunitary symmetries $\theta\{C_{2b}|0\}$, $\theta\{\sigma_{da}|0\}$. Again, none of these symmetries would reverse the spin up (down) state to its opposite, therefore spin splitting is expected in this case.

The situation becomes a bit more complicated for k-path on the boundary of BZ. Along Z-R with $\bm{k} = (u, 0, 1/2)$, the possible symmetries that keep $\bm{k}$ invariant are:

$$\begin{aligned}
\{E|0\} : (u, 0, 1/2) &\rightarrow (u, 0, 1/2) \\
\{\sigma_h|0\}: (u, 0, 1/2) &\rightarrow (u, 0, 1/2) - (0, 0, 1) \\
\{C_{2x}|\bm{\tau}\}: \left(u, 0, \frac{1}{2}\right) &\rightarrow \left(u, 0, \frac{1}{2}\right) - (0, 0, 1) \\
\{\sigma_{vy}|\bm{\tau}\}: (u, 0, 1/2) &\rightarrow (u, 0, 1/2) \\
\theta\{C_{2y}|\bm{\tau}\}: (u, 0, 1/2) &\rightarrow (u, 0, 1/2) \\
\theta\{\sigma_{vx}|\bm{\tau}\}: (u, 0, 1/2) &\rightarrow (u, 0, 1/2) - (0, 0, 1)
\end{aligned} \quad (S4)$$

and their combination with a primitive translation along z axis of $\{E|(0,0,1)\}$.

(1) When SOC is ignored, the group of the wavevector are composed of $\{E|0\}$, $\{\sigma_h|0\}$, $\theta\{C_{2y}|\bm{\tau}\}$ and $\theta\{\sigma_{vx}|\bm{\tau}\}$ symmetries and their combination with $\{E|(0,0,1)\}$, where either $\theta\{C_{2y}|\bm{\tau}\}$ or $\theta\{\sigma_{vx}|\bm{\tau}\}$ will protect a double degeneracy.

(2) When SOC is considered, the group of the wavevector have only unitary symmetries $\{E|0\}$, $\{\sigma_h|0\}$, $\{C_{2x}|\bm{\tau}\}$, and $\{\sigma_{vy}|\bm{\tau}\}$ and their combination with $\{E|(0,0,1)\}$. $\{E|0\}$ and $\{\sigma_h|0\}$ both are unit $2 \times 2$ matrix in the spin space hence will not introduce any spin degeneracy; while by selecting basis as spin polarization along $y$, neither $\{C_{2x}|\bm{\tau}\}$ nor $\{\sigma_{vy}|\bm{\tau}\}$ will reverse the spin polarization. As the consequence, one would expect spin splitting along Z-R when SOC is included.

## APPENDIX E: Power of k dependence of spin splitting in AFM MnF$_2$ with SOC

We have calculated the scaling of the spin splitting with wave vector from DFT calculation: Nearby Γ (progressing along Γ-M), the splitting between the V1 and V2 bands shows a quadratic relation to wavevector as $\bm{k}^\eta$ with a numerically fitted value of $\eta = 1.98$, while nearby Z (progressing along Z-R), such splitting shows a linear relation as $\bm{k}^\eta$ with numerical $\eta = 0.98$, ; the same quadratic and linear relations also hold for the splitting between V3 and V4. (see Table SI)

**Table SIII | Power of k dependence of spin splitting in AFM MnF$_2$ with SOC.** The spin splitting near given high symmetry $\bm{k}_0$ point and $(\bm{k}_0 + \Delta\bm{k})$ are fitted to $\alpha_{\bm{k}_0}|\bm{k}|^\eta$ for the top two valence bands V1, V2 and the third and fourth valence bands V3, V4. Row captions like $\bm{\Gamma_{\Gamma-M}}$ are used to note spin splitting near Γ along Γ-M.

| k point | $\Delta_{SS}^{V1-V2}(k)$ | $\Delta_{SS}^{V3-V4}(k)$ | Linear or Quadratic? |
|---|---|---|---|
| $\Gamma_{\Gamma-M}$ | $0.45k^{1.98}$ | $3.72k^{1.95}$ | Quadratic |
| $M_{M-\Gamma}$ | $0.44k^{1.98}$ | $3.86k^{1.95}$ | Quadratic |
| $Z_{Z-R}$ | $0.04k^{0.98}$ | $0.02k^{1.01}$ | Linear |
| $A_{A-R}$ | $0.01k^{1.04}$ | $0.05k^{0.98}$ | Linear |

**APPENDIX F: EFFECTIVE TWO-BAND MODEL HAMILTONIAN AT SPECIFIC $k$ POINTS MODEL IN AFM MNF2**

In AFM MnF$_2$, one can define two spin-related AFM local atomic basis states[24] with one spin-up state localized mostly on Mn1 at (0, 0, 0) and one spin-down state mostly localized on Mn2 at (1/2, 1/2, 1/2). The AFM ordering is thus embedded in the inequivalence distribution on Mn1 and Mn2 of the spin-related basis. The effective two-band model Hamiltonian at specific $k$ points can then be determined by the constrains imposed by the symmetries of the group of wavevector on the basis.

**6.1 Effective model at $\Gamma$:** At the $\Gamma$ point, the group of wavevector inherits all the symmetries that the AFM magnetic space group has. Upon applying the symmetries (only symmetry generators are needed) on the AFM basis one can find the representations and transformation properties of the Pauli matrix $\boldsymbol{\sigma}$ and tensor operator $\boldsymbol{k}$.

(1) When the SOC is ignored, the spin orientation is enforced to align along the magnetization direction, i.e., z-axis. The corresponding magnetic space group is P4$_2$'/mn'm, which can be generated by three unitary symmetries $\{C_{2a}|0\}, \{C_{2b}|0\}, \{I|0\}$ and one antiunitary symmetry $\theta\{C_{2x}|\boldsymbol{\tau}\}$. Table SIV lists the transformation properties of the Pauli matrix $\boldsymbol{\sigma}$ and tensor operator $\boldsymbol{k}$ under these symmetry operations, the only possible invariant spin splitting term that could exists in Hamiltonian is $\sigma_z k_x k_y$, indicating quadratic dependence of spin splitting to displacement in $\boldsymbol{k}$ along the diagonal $\Gamma$-M direction, and spin degeneracy along $\Gamma$-X direction in agreement with our DFT results seen in Figure 2(b).

**Table SIV | The transformation properties of symmetrized matrix and irreducible tensor up to the second order in $k$ under symmetry operations of the group of wavevector at $\Gamma$ (without SOC).**

| Symmetrized matrix | Irreducible tensor | $\{C_{2a}|0\}$ | $\{C_{2b}|0\}$ | $\{I|0\}$ | $\theta\{C_{2x}|\boldsymbol{\tau}\}$ |
|---|---|---|---|---|---|
| $\sigma_0$ | $C, k_x^2 + k_y^2, k_z^2$ | 1 | 1 | 1 | 1 |
| $\sigma_z$ | $k_x k_y$ | 1 | 1 | 1 | -1 |

(2) When including SOC, the corresponding magnetic space group becomes P4$_2$'/mnm', which can be generated by three unitary symmetries $\{C_{2x}|\boldsymbol{\tau}\}, \{C_{2y}|\boldsymbol{\tau}\}, \{I|0\}$ and one antiunitary symmetry $\theta\{C_{2a}|0\}$. The transformation properties of the Pauli matrix $\boldsymbol{\sigma}$ and tensor operator $\boldsymbol{k}$ are listed in Table SV:

**Table SV | The transformation properties of symmetrized matrix and irreducible tensor up to the second order in $k$ under symmetry operations of the little point group at $\Gamma$ (with SOC)**

| Symmetrized matrix | Irreducible tensor | $\{C_{2x}\|\boldsymbol{\tau}\}$ | $\{C_{2y}\|\boldsymbol{\tau}\}$ | $\{I\|0\}$ | $\theta\{C_{2a}\|0\}$ |
|---|---|---|---|---|---|
| $\sigma_0$ | $C, k_x^2 + k_y^2, k_z^2$ | 1 | 1 | 1 | 1 |
| $\sigma_z$ | $k_x k_y$ | -1 | -1 | 1 | 1 |
| $(\sigma_x, \sigma_y)$ | - | $\begin{bmatrix} 1 & 0 \\ 0 & -1 \end{bmatrix}$ | $\begin{bmatrix} -1 & 0 \\ 0 & 1 \end{bmatrix}$ | $\begin{bmatrix} 1 & 0 \\ 0 & 1 \end{bmatrix}$ | $\begin{bmatrix} 0 & -1 \\ -1 & 0 \end{bmatrix}$ |

We see from Table SV that the only possible invariant spin splitting term up to second order in $\boldsymbol{k}$ is $\sigma_z k_x k_y$, indicating quadratic dependence of spin splitting on variations in $\boldsymbol{k}$ along the diagonal Γ-M direction when SOC is included. (see DFT results in Figure 1(c)) The effective Hamiltonian term $\sigma_z k_x k_y$ also captures the four-quadrant pattern of the out-of-plane spin polarization as $k_x k_y$ having opposite signs in first and third quadrants and in second and fourth quadrants as seen in Figure 3.

### 6.2 Effective model at Λ

(1) When SOC is excluded, the spin splitting term takes exactly the form as at Γ, $\sigma_z k_x k_y$. So no spin splitting along Λ-W and quadratic spin splitting along Λ-V.

(2) When SOC is included, from Table SVI, the effective spin splitting terms are linear combinations of $H_\Lambda = A\sigma_z k_x k_y + B(\sigma_x k_y + \sigma_y k_x)$, here A and B are real coefficients. One would then also expect quadratic dependence along the diagonal Λ-V direction from the first term and linear dependence along Λ-W direction from the second term. Also, the four quadrants pattern and Dresselhaus-like spin texture can be explicitly captured by the effective model Hamiltonian, the first term contributes to the four quadrants pattern with out-of-plane spin polarization, where $k_x k_y$ takes has opposite sign in first ($k_x > 0, k_y > 0$) and third ($k_x < 0, k_y < 0$) quadrants relative to the second ($k_x < 0, k_y > 0$) and fourth ($k_x > 0, k_y < 0$) quadrants. The second term resembles in analytical form to conventional Dresselhaus term and contributes the in-plane Dresselhaus spin texture. One should note that it is the distribution of the spin-related basis on two Mn that provides non-vanishing in-plane spin texture. If the spin up and spin down basis localized completely on Mn1 and Mn2, the in-plane spin texture would vanish and there would be no Dresselhaus spin texture at all. Larger mixing between the two local Mn atoms of the spin basis will lead to a stronger in-plane spin polarization.

**Table SVI | The transformation properties of symmetrized matrix and irreducible tensor under symmetry operations of the little point group at Λ with SOC.**

| Symmetrized matrix | Irreducible tensor | $\{\sigma_{vx}\|\boldsymbol{\tau}\}$ | $\{\sigma_{vy}\|\boldsymbol{\tau}\}$ | $\theta\{C_{2a}\|0\}$ |
|---|---|---|---|---|
| $\sigma_0$ | $C, k_x^2 + k_y^2, k_z^2$ | 1 | 1 | 1 |
| $\sigma_z$ | $k_x k_y$ | -1 | -1 | 1 |
| $(\sigma_x, \sigma_y)$ | $(k_y, k_x)$ | $\begin{bmatrix} 1 & 0 \\ 0 & -1 \end{bmatrix}$ | $\begin{bmatrix} -1 & 0 \\ 0 & 1 \end{bmatrix}$ | $\begin{bmatrix} 0 & -1 \\ -1 & 0 \end{bmatrix}$ |

### 6.3 Effective model at $Z$

(1) When SOC is excluded, the spin splitting term takes exactly the form as at $\Gamma$, $\sigma_z k_x k_y$ and give rise to zero spin splitting along Z-R, and quadratic-in-k spin splitting along Z-A. (see DFT band structure in Figure 2(b)).

(2) When SOC is included, from Table SVII, the spin splitting terms are linear combinations of $H_Z = A\sigma_z k_x k_y + B(\sigma_x k_y - \sigma_y k_x)$, here A and B are real coefficients. Once again, the spin splitting will have quadratic dependence along the diagonal Z-A direction and a four-quadrants out-of-plane spin polarization pattern from the first term, and linear dependence along the Z-R direction from the second term. Moreover, despite the fact that the second term resembles in form the conventional Rashba Hamiltonain[2], it will not create in-plane spin polarization (see Figure S1). The vanishing spin polarization is the consequence of zero mixing between the two spin-related Bloch basis of Mn1 and Mn2 for the same spin, that are $|Mn_1, \uparrow\rangle$ and $|Mn_2, \uparrow\rangle$ (also $|Mn_1, \downarrow\rangle$ and $|Mn_2, \downarrow\rangle$), which form a pair of zero in-plane spin polarized but non-zero splitting states, $a|Mn_1, \uparrow\rangle + b|Mn_2, \downarrow\rangle$ and $a|Mn_1, \uparrow\rangle - b|Mn_2, \downarrow\rangle$ with a, b being the complex constant coefficients of the states satisfying the normalization condition $|a|^2 + |b|^2 = 1$. When $a = b$, the out-of-plane spin polarization also vanishes. The zero mixing between $|Mn_1, \uparrow\rangle$ and $|Mn_2, \uparrow\rangle$ (and between $|Mn_1, \downarrow\rangle$ and $|Mn_2, \downarrow\rangle$) is enforced by $\{\sigma_h|0\}$ symmetry at Z, since $|Mn_1, \uparrow\rangle$ (and $|Mn_2, \downarrow\rangle$) takes opposite eigenvalue of $\{\sigma_h|0\}$ symmetry to $|Mn_2, \uparrow\rangle$ (and $|Mn_1, \downarrow\rangle$). The same reason accounts for the zero in-plane spin polarization but non-zero splitting observed at R when SOC is included, as $\{\sigma_h|0\}$ also being a symmetry of $\boldsymbol{k}$ at R and forbids mixing between $|Mn_1, \uparrow\rangle$ and $|Mn_2, \uparrow\rangle$ (and between $|Mn_1, \downarrow\rangle$ and $|Mn_2, \downarrow\rangle$). The surprising effect of spin splitting with vanishing spin polarization was also reported recently in non-magnetic crystals (e.g., bulk GaAs[56] and graphene[57]).

**Table SVII | The transformation properties of symmetrized matrix and irreducible tensor under symmetry operations of the little point group at Z with SOC.**

| Symmetrized matrix | Irreducible tensor | $\{C_{2x}|\boldsymbol{\tau}\}$ | $\{C_{2y}|\boldsymbol{\tau}\}$ | $\{I|0\}$ | $\theta\{C_{2a}|0\}$ |
|---|---|---|---|---|---|
| $\sigma_0$ | $C, k_x^2 + k_y^2, k_z^2$ | 1 | 1 | 1 | 1 |
| $\sigma_z$ | $k_x k_y$ | -1 | -1 | 1 | 1 |
| $(\sigma_x, \sigma_y)$ | $(k_y, -k_x)$ | $\begin{bmatrix} -1 & 0 \\ 0 & 1 \end{bmatrix}$ | $\begin{bmatrix} 1 & 0 \\ 0 & -1 \end{bmatrix}$ | $\begin{bmatrix} -1 & 0 \\ 0 & -1 \end{bmatrix}$ | $\begin{bmatrix} 0 & 1 \\ 1 & 0 \end{bmatrix}$ |

**APPENDIX G: In-plane Spin texture in MnF$_2$ on the $\boldsymbol{k}$-planes $k_z = 0$ and $k_z = \pi/c$**

Figure S1 shows the calculated spin textures of the V1 and V2 bands in MnF$_2$ on $\boldsymbol{k}$-planes $k_z = 0$ and $k_z = \pi/c$, where c is the lattice constant along (001). For the out-of-plane spin polarization, we find the same four-quadrant pattern as the one found on $k_z = \pi/2c$ plane (shown in Figure 3). While, in contrast to in-plane Dresselhaus spin texture observed on $k_z = \pi/2c$ plane, on the $\boldsymbol{k}$-planes $k_z = 0$ and $k_z = \pi/c$, there is no in-plane spin polarization.

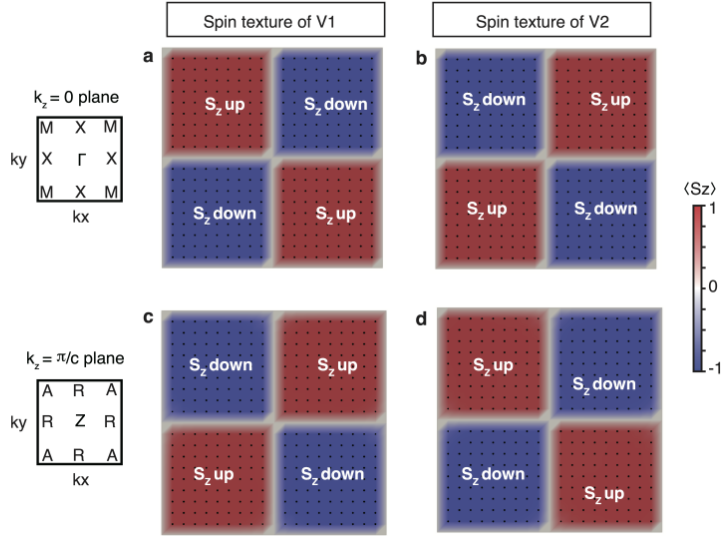

**Figure S1 | Spin textures of the top two valence bands (V1 and V2) in AFM MnF$_2$ on two $k_z$ planes: (a) (b)** $k_z = 0$ plane, and **(d) (e)** $k_z = \pi/c$ plane. For each $k_z$ plane, the labels of the high symmetry $k$ points are shown by a diagram on the left side of each horizonal panel. The in-plane spin polarizations are indicated by black arrows, while black dot means the in-plane polarization at this $k$ is zero; the out-of-plane spin polarizations are mapped by colors from blue to red.

**APPENDIX H: The spin splitting induced by cooperative effects of AFM and SOC in MnF$_2$**

Allowing SOC in a NM model does not lead to any spin splitting since the $\theta I$ symmetry is always preserved (see Figure S2 (a)). In contrast, introducing SOC to AFM leads to cooperative effects of AFM + SOC. For example, it creates additional spin splitting along certain $k$-paths, e.g., Z-R, R-A and X-R directions (see Figure S2(b)). This is because in the AFM phase described without SOC, the spin degeneracy along Z-R, R-A and X-R directions is guaranteed by the symmetry operations $\theta\{C_{2x}|\tau\}$, $\theta\{C_{2y}|\tau\}$, $\theta\{\sigma_{2x}|\tau\}$, and $\theta\{\sigma_{2y}|\tau\}$; adding SOC to pre-existing AFM couples the real space rotations to spin operations and breaks all four anti-unitary symmetries, leading therefore to spin splitting along these directions (see full details about how SOC induces spin splitting in preexisting AFM along Z-R in Appendix E). An interesting fact is that we find spin splitting at the R point (which is TRIM) when adding SOC to AFM phase. This manifests the breaking of time reversal symmetry in the AFM phase. The lifting of spin degeneracy at TRIM point and its connected $k$ paths represents a cooperative effect of magnetism and SOC: neither AFM without SOC (Figure 2(b)) nor SOC without AFM (Figure S2(a)) shows spin splitting along these directions, but the coexistence of SOC and AFM leads to spin splitting.

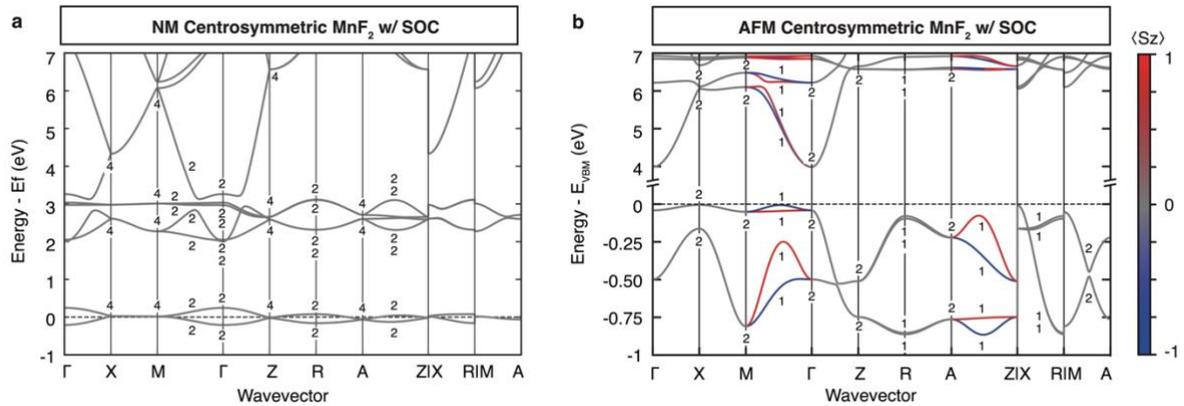

**Figure S2 | DFT band structures of centrosymmetric MnF$_2$ in NM and AFM with SOC.** In all cases we use the experimentally observed centrosymmetric tetragonal structure[34]: (a) NM with SOC; (b) AFM with SOC. Out-of-plane spin polarizations are mapped to color scales from blue to red. The integer numbers attached to bands are degeneracy factors.